\documentclass[aip,amsmath,amssymb,reprint]{revtex4-1}
\usepackage{graphicx,bm,times,url,dcolumn}
\usepackage{amsmath}
\usepackage{amsfonts}
\usepackage{siunitx}
\usepackage{epstopdf}
\usepackage{natbib}
\usepackage{verbatim}
\graphicspath{{./fig/}{./png/}}

\newcommand{\EQ}{\begin{equation}}
\newcommand{\EN}{\end{equation}}
\newcommand{\EQA}{\begin{eqnarray}}
\newcommand{\ENA}{\end{eqnarray}}

\newcommand{\Eqss}[2]{Equations~(\ref{#1})--(\ref{#2})}

\newcommand{\App}[1]{Appendix~\ref{#1}}

\newcommand{\Fig}[1]{Figure~\ref{#1}}

{}
{}
{}

{}
{}
{}
{}
{}
{}
{}
{}
{}
{}
{}
{}
{}
{}
{}
{}
{}
{}
{}
{}
{}

{}
{}
{}

{}

{}
{}

\newcommand{\BB}{\bm{B}}

\newcommand{\uu}{\mbox{\boldmath $u$} {}}
\newcommand{\UU}{\mbox{\boldmath $U$} {}}

\newcommand{\SSS}{\mbox{\boldmath $S$} {}}
\newcommand{\AAA}{\mbox{\boldmath $A$} {}}

\newcommand{\ee}{\mbox{\boldmath $e$} {}}
\newcommand{\nn}{\mbox{\boldmath $n$} {}}

\newcommand{\FF}{\mbox{\boldmath $F$} {}}

\newcommand{\nab}{\mbox{\boldmath $\nabla$} {}}
\newcommand{\OO}{\bm{\Omega}}

\newcommand{\oo}{\mbox{\boldmath $\omega$} {}}

\newcommand{\DD}{{\rm D} {}}

\newcommand{\dd}{{\rm d} {}}

\def\Ro{\mbox{\rm Ro}}

\def\Rey{\mbox{\rm Re}}

\def\cs{c_{\rm s}}

\usepackage{color}
\usepackage[utf8]{inputenc}
\usepackage[T1]{fontenc}
\usepackage{mathptmx}

\newcommand{\RN}[1]{%
  \textup{\uppercase\expandafter{\romannumeral#1}}%
}

\begin{document}

\preprint{AIP/123-QED}

\title{Topological Constraints in the Reconnection of Vortex Braids}

\author{S. Candelaresi}
\affiliation{School of Mathematics and Statistics, University of Glasgow, Glasgow G12 8QQ, United Kingdom}
\affiliation{Division of Mathematics, University of Dundee, Dundee DD1 4HN, United Kingdom}
\author{G. Hornig}
\affiliation{Division of Mathematics, University of Dundee, Dundee DD1 4HN, United Kingdom}
\author{B. Podger}
\affiliation{Division of Mathematics, University of Dundee, Dundee DD1 4HN, United Kingdom}
\author{D. I. Pontin}
\affiliation{School of Mathematical and Physical Sciences, University of Newcastle, Callaghan, NSW 2308, Australia}
\affiliation{Division of Mathematics, University of Dundee, Dundee DD1 4HN, United Kingdom}

\date{\today}

\begin{abstract}
We study the relaxation of a topologically non-trivial vortex braid with
zero net helicity in a barotropic fluid.
The aim is to investigate the extent to which the topology of the vorticity field -- characterized by
braided vorticity field lines -- determines the dynamics, particularly the asymptotic behaviour under
vortex reconnection in an evolution at high Reynolds numbers ($25,000$).
Analogous to the evolution of braided magnetic fields in plasma, we find that the relaxation of our vortex braid leads to
a simplification of the topology into large-scale regions of opposite swirl, consistent with an inverse cascade of the helicity.
The change of topology is facilitated by a cascade of vortex reconnection events.
During this process the existence of regions of positive and negative kinetic helicity imposes a lower
bound for the kinetic energy.
For the enstrophy we derive analytically a lower bound given by the presence of unsigned
kinetic helicity, which we confirm in our numerical experiments.
\end{abstract}

\maketitle

\section{Introduction}

It is well established that the degree of tangling/knottedness of vorticity field lines can have important implications
for the dynamics of a fluid \citep{MoffattKnottedness1969,moffatt1992}.
In a barotropic fluid in the ideal case with zero viscosity this tangling is preserved,
restricting the lowest energy state to which the fluid has access.
This has been demonstrated both  in experiments \citep{Scheeler-Kleckner-2014-11-43-PNAS, Scheeler-Rees-2017-357-487-Sci}
and numerical simulations \citep{kerr_2018}.
When the Reynolds number is large but finite, vortex reconnection may take place, permitting a change of topology of the vortex lines.
Individual events of such vortex reconnection have been studied, typically involving reconnection between isolated vortex
tubes or rings \citep{ashurst1987, melander1989, kida1991b, vanrees2012, mcgavin2018, Zhao-Yu-2021-910-A31-JFluidMech}.
Notably, in these simulations the vortex tubes usually contort during their mutual approach, such that the vortex lines reconnect
locally anti-parallel (in a 2d plane).
However, in many applications, the vorticity is a smooth non-vanishing function across the
volume and can't be modelled as a set of interacting isolated tubes.
Examples include rotating stars and planets where there is a dominant
direction of the vorticity (aligned with the rotation axis) onto which contributions from
convection are superimposed.
The resulting field could be interpreted as a vortex braid.
If vortex reconnection  occurs in such a scenario, the presence of a dominant uni-directional
vorticity field means that the reconnection is fully three-dimensional \citep{Hornig:2001wx}, as recently
observed in the reconnection of vortex tubes with swirl \citep{vanrees2012, mcgavin2019}.
Note that with ``vortex reconnection'' here we refer to the process by which the vorticity field lines change their topology,
prohibited in a barotropic, inviscid fluid.
This should not be confused with the notion of reconnection of vorticity isosurfaces, also sometimes referred to as vortex reconnection.

We analyze in the following the relaxation of a braided vorticity field in a fluid of high Reynolds number ($\Rey > 10^4$).
The  aim  is  to  investigate  the  extent  to  which  the  topology  of the vorticity field -- characterized by braided vorticity
field lines -- determines the dynamics.
With the notion ``braided'' we describe a situation where we have a dominant component of the vorticity
field in one direction, in our case the $z$-direction, so that all vorticity field lines
connect two opposite sides of our domain (see \Fig{fig: initial_setup}).
The motivation for this scenario is three-fold: First, the situation of a braided vorticity field is of relevance
for many cases of rotating astrophysical bodies, where the rotation of the star or planet provides a dominant
component of the vorticity and the contributions from convection or turbulence to the vorticity are weaker and
only contribute to a braiding of the vortex lines.
Second, this set-up has the advantage that all vorticity lines connect from the lower to upper boundary of our domain.
That is, there are no null points of the vorticity in the domain and hence the topological structure of the field
is uniquely described by its vorticity-field line mapping from the lower to the upper boundary \citep{Yeates:2013gi}.
This allows us to analyze the topology of the field at any point in time using various tools such as the field
line helicity \citep{Yeates-Hornig-2011-18-102118-PhysPlasm,Yeates:2013gi}, the topological entropy
\citep{Candelaresi-Pontin-2017-27-9-Chaos}, or the topological degree \citep{Yeates_Topology_2010}.
With these tools one can follow the dynamics of the relaxation with the ultimate aim to make predictions
regarding the final state of the relaxation process.
One can even identify individual processes of vortex reconnection taking place.
However, in this study we are less interested in the individual reconnection events
and more in the collective effect
that a turbulent cascade of reconnection events has on the route the relaxation process takes.
The third motivation is that this vortex braid relaxation is the exact analogue to a magnetic braid
relaxation studied by the authors before \citep{Wilmot-Smith-Hornig-2009-696-1339-ApJ}.
In these previous studies the relaxation exhibited additional constraints on the dynamics,
over and above the one imposed by the conservation of helicity \cite{Yeates_Topology_2010}.
To investigate the presence of such constraints in vortex dynamics is the aim of this study.

\section{Model}

\subsection{Initial Condition}

\begin{figure}\begin{center}
\includegraphics[width=0.4\columnwidth]{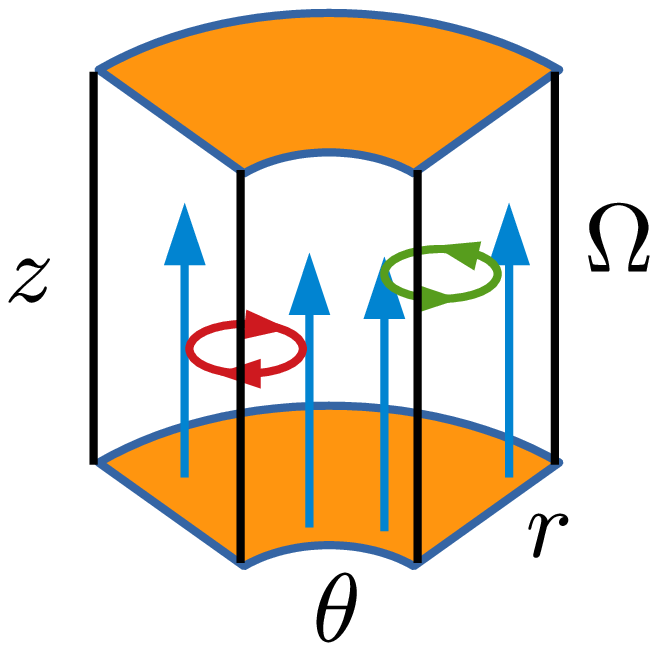}
\includegraphics[width=0.4\columnwidth]{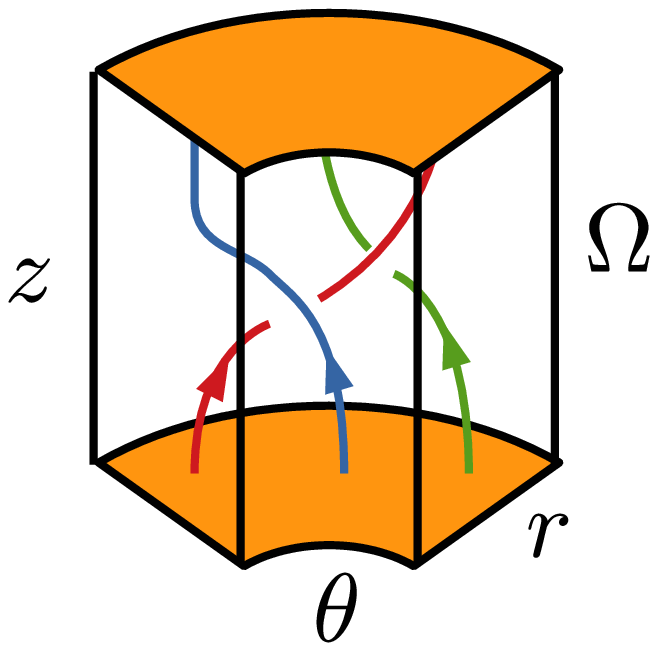}
\end{center}
\caption[]{
Schematic representation of the initial set up of the vortex braid.
Left we represent the components of the field
with the two vortex rings (red and green) of opposing vorticity sign and the background vorticity (blue).
On the Right we show the superposition of the vortex rings with the background field
which results in the vortex braid.
}
\label{fig: initial_setup}\end{figure}

We wish to construct a vortex braid 
in which all vorticity field lines connect between opposite boundaries of the domain.
In order to facilitate direct comparison with a well-studied magnetic braid
we choose the particular braiding pattern of the vorticity lines to be analogous to that of the magnetic field lines in that magnetic braid
\citep{Wilmot-Smith-Hornig-2009-696-1339-ApJ}.
The vorticity field consists of a constant background field in the $z$-direction together with two vortex rings with
their symmetry axes also aligned to the $z$-direction.
The vortex rings are located such that we obtain vorticity field lines as in \Fig{fig: initial_setup}
(see also \Fig{fig: streamlines_t5750}).
All vorticity lines connect between opposite (plane-parallel, constant-$z$) boundaries.
The background vorticity field is conveniently obtained from a (solid-body) rotational flow with the $z$-axis as the axis of rotation.
An illustration is shown in \Fig{fig: initial_setup}, which corresponds to the vortex field we will be using.

To avoid complications of generation of secondary vortices in domain corners, something that
typically occurs in Cartesian geometries for rotating fluids, we make use of a cylindrical domain,
which rotates about the $z$-axis.
We construct the field first without the homogeneous background
in a cylindrical wedge with periodic azimuthal boundaries and move the frame of reference
together with the global rotational flow.
To obtain the effect of the background field (and the full vortex braid) we add a Coriolis term in
the momentum equations (see below).

Each of the vortex rings (red and green in \Fig{fig: initial_setup}) 
is constructed by first defining a single vortex ring centered at the origin.
This greatly simplifies our calculations, due to the ring's azimuthal symmetry.
We then translate the calculated field to its position in the wedge domain;
a non-trivial transformation, as described in \App{sec: construction of vortex tubes}.

For our computational domain we choose a cylindrical wedge of dimensions
$r \in [45, 65]$, $\theta \in [-0.1,0.1]$ and $z \in [-16, 16]$.
We choose the $\theta$ and $z$ directions to be periodic, while the boundary conditions in the radial
direction are chosen such that the normal component of the velocity vanishes
and any mass flux is suppressed.

Within this domain we place two vortex rings of opposite orientation, with axes lying in planes of constant $z$ and centers
at positions $(r, \theta, z) = (55, \arctan{(1/55)}, -8)$ and $(r, \theta, z) = (55, -\arctan{(1/55)}, 8)$.
This means that the subsection of our volume in which the vortex lines exhibit a non-trivial tangling is located centrally
within the domain, away from the $r$ and $\theta$ boundaries.
The initial vertical distance of 16 in non-dimensional code units between the (axes of the) vortex rings ensures that
the velocities induced by the two vortex rings do not significantly overlap at $t=0$.
Note that the superposition of the vortex ring with the background vorticity leads to a local twisting of the vorticity lines,
and since the boundaries are periodic along $z$, the vorticity lines in principle pass through infinitely many of these vortex rings.

To prevent effects from supersonic flows we choose the amplitude of the vorticity in the vortex
rings to $\alpha = 0.1$.
This will keep the velocities throughout the simulations well below the sound speed of $1$.
For the background vorticity we choose $\OO = 0.1\ee_z$.
This will lead to a vorticity field with the desired topology.
The ratio of the two amplitudes $\alpha/\Omega$ determines the strength of the braiding
and with that the topology of the vortex field.
Note that this constant background vorticity refers to the rest frame, and is achieved by using a Coriolis term in
the simulations with the angular velocity $\tilde{\OO} = \OO/2$.

With these parameters we obtain a Rossby number
\EQ
\Ro = \frac{u}{2L\tilde{\Omega}},
\EN
where $u$ is a typical velocity and $L$ a typical length scale.
In our case $u \approx 0.1$ (velocity at the vortex rings),
$L \approx 1$ (size of the vortex rings)
and $\tilde{\Omega} = 0.05$.
With that we have $\Ro \approx 1$.

\subsection{Numerical Setup}

As described above,
to circumvent issues at the domain's corners and issues with non-vanishing normal velocities
at the boundaries, we place our cylindrical wedge domain in a co-moving frame.
This generates the additional term of the Coriolis force $2\uu\times\tilde{\OO}$.
Our resulting equations are then the equations of motion for a viscous, isothermal and compressible gas:
\EQ
\frac{\DD \uu}{\DD t} = -\cs^{2} \nab \ln{\rho} + 2\uu\times\tilde{\OO} + \FF_{\rm visc},
\label{eq: dUdt}
\EN
\EQ
\frac{\DD \ln{\rho}}{\DD t} = -\nab \cdot \uu,
\label{eq: drhodt}
\EN
with the isothermal speed of sound $\cs$, density $\rho$,
viscous forces $\FF_{\rm visc}$ and Lagrangian time derivative
$\DD/\DD t = \partial/\partial t + \uu\cdot\nab$.
Here the viscous forces are given as $\FF_{\rm visc} = \rho^{-1}\nab\cdot2\nu\rho\SSS$,
with the kinematic viscosity $\nu$, and traceless rate of strain tensor
$\SSS_{ij} = \frac{1}{2}(u_{i,j}+u_{j,i}) - \frac{1}{3}\delta_{ij}\nab\cdot\uu$.
Being an isothermal gas we have $p = \cs^{2}\rho$ for the pressure. 
Note that since $\cs^2$ is constant, $\nabla p\times\nabla \rho=0$, meaning that there is no baroclinic vorticity production.

\Eqss{eq: dUdt}{eq: drhodt} are solved using the {\sc PencilCode}
\footnote{github.com/pencil-code/}, which is an Eulerian finite-difference code using sixth-order spatial derivatives
and a third-order time-stepping scheme \citep{BD02PC}.
Throughout our simulations we use $\nu = 10^{-3}$ to $\nu = 4\times 10^{-5}$ in order to reduce
kinetic energy dissipation and kinetic helicity dissipation as much as the limited resolution
of $512\times256\times256$ ($r$, $\theta$, $z$) grid points allows.
We emphasize that due to the barotropic nature of the fluid, in the inviscid case the tangling (or braiding)
of the vortex lines would be preserved for all time.

\subsection{Incompressibility}

By construction the initial velocity field has the property $\nab\cdot\uu \approx 0$.
Being approximately incompressible, any calculations involving the evolution of the kinetic energy
or enstrophy significantly simplify.
This implies that the initial uniform density does not change in time
(see equation \eqref{eq: drhodt}).
However, numerical errors in the calculation of the potential $C_0$ (equation \eqref{eq: C0 potential})
can cause deviations from $\nab\cdot\uu = 0$.

\begin{figure}\begin{center}
\includegraphics[width=0.95\columnwidth]{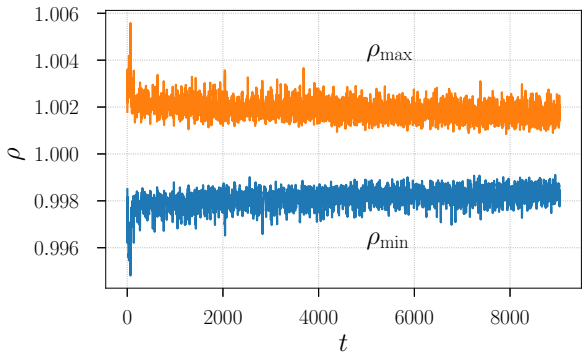}
\end{center}
\caption[]{
Minimum and maximum density in the domain as a function of time
for the simulation with viscosity $\nu = 4\times 10^{-5}$.
}
\label{fig: rho_min_max}\end{figure}

To check if our assumption of incompressibility holds true for all time
we plot the maximum and minimum density in the domain as a function of time
(\Fig{fig: rho_min_max}).
We observe a deviation of ca.\ $0.5$\% from the uniform density at initial time, which quickly
decreases to ca.\ $0.2$\% and approximately remains constant.
Being such a small deviation we can safely assume that the system is approximately incompressible.

\section{Evolution of the System}

\begin{figure}\begin{center}
\includegraphics[width=0.95\columnwidth]{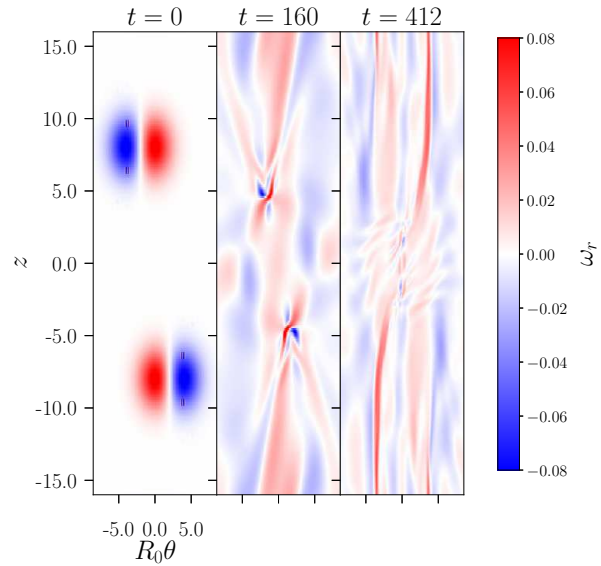}
\end{center}
\caption[]{
Slices through the domain for $\omega_r$ at different times at $r = R_0 = 55$ for the simulation with
viscosity $\nu = 4\times 10^{-5}$.
}
\label{fig: omega_r slices}\end{figure}

Following initiation of the simulation, the two vortex rings travel towards one another due to their self-induced motion,
meeting approximately at the mid-plane, $z = 0$.
This is analogous to the self-induced motion of an isolated infinitesimal vortex ring
(i.e.~with infinitesimal minor radius), with some distortion due to the
presence of the background vorticity and the finite radius of the rings.
Due to the offset in $\theta$ between the two rings, they do not meet face-on
(\Fig{fig: omega_r slices}).
However, their collision leads to a local enhancement of the vorticity where they meet, as seen in the
enstrophy evolution (see section \ref{sec: enstrophy}).

From previous studies of the relaxation of magnetic braids we know that the braid constructed in this way
requires many reconnection events to untangle and is very efficient in generating a turbulent evolution.
Following the initial collision, we see indeed that a highly-fluctuating, ``turbulent-like'' evolution ensues,
in which we find numerous locations at which vortex reconnection takes place
(identified by calculating $(\nabla\times\oo)\cdot(\oo+\OO)/|\oo+\OO|$, see \citep{Hornig:2001wx, mcgavin2018}).
Through these many localized reconnection events the field topology simplifies, with the vortex lines becoming less tangled.
However, the final state retains a non-trivial topology, and our main purpose is to analyse the way in
which this final state is determined by the
initial field topology.

\section{Enstrophy}
\label{sec: enstrophy}

Unlike the total energy and the kinetic helicity, the enstrophy
\EQ \label{eq; enstrophy}
\mathcal{E} = \int_{V}\oo^2\ \dd V
\EN
is not necessarily conserved, even in the inviscid case.
To see the factors that can lead to a change in enstrophy, we
use the momentum equation \eqref{eq: dUdt} to write the vorticity equation as
\EQA 
\frac{\partial \oo}{\partial t} & = & \nab\times(\uu\times\oo) + 2\nab\times(\uu\times\tilde{\OO}) + \nu\Delta\oo \nonumber \\
 & & + 2\nu\nab\times(\nab\ln(\rho)\cdot\SSS). \label{eq: vorticity}
\ENA
With this we can write the time evolution of the total enstrophy as
\EQA
\frac{\dd \mathcal{E}}{\dd t} & = & 2\int_{V} \oo\cdot\frac{\partial\oo}{\partial t}\ \dd V \nonumber \\
 & = & 2\int_{\partial V} ((\uu\cdot\oo)\oo + \nu\oo\times\nab\times\oo) \cdot\dd\SSS \nonumber \\
 & & + 2\int_V \left( (\uu\times(\oo+2\tilde{\OO}))\cdot\nab\times\oo 
     - \nu(\nab\times\oo)^2 \right. \nonumber \\
 & & \left. + 2\nu\oo\cdot\nab\times(\nab\ln(\rho)\cdot\SSS) \right) \ \dd V \label{eq: enstrophy},
\ENA
where we used the fact that the azimuthal and vertical dimensions are periodic,
$\uu\cdot\nn = 0$ at the $r$ boundaries and $\tilde{\OO}\cdot\nn = 0$ on the $r$ and $\theta$ boundaries.

Apart from the terms involving viscosity, we have two more volume terms and one surface term that in general do not vanish.
This is interesting, since our domain is closed in the radial direction and yet, there can
be enstrophy fluxes through those boundaries.
However, throughout all of our simulations the velocities near the radial boundaries are very small
and this term can be safely ignored.
The first volume term describes the dynamical generation or annihilation of enstrophy according
to the alignment of the velocity, vorticity and its curl, while the second volume term
describes the dynamical generation/annihilation of enstrophy due to the Coriolis force.

\begin{figure}\begin{center}
\includegraphics[width=0.95\columnwidth]{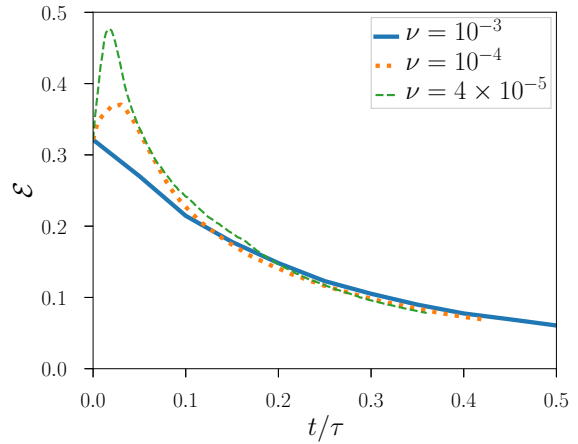}
\end{center}
\caption[]{
Enstrophy evolution for different viscosities $\nu$
against normalized (diffusion) time (see equation \eqref{eq: dissipation time}).
}
\label{fig: enstrophy_compare}\end{figure}

For high Reynolds numbers we observe first an increase and then a gradual decrease in enstrophy (\Fig{fig: enstrophy_compare}).
As the vortex rings approach and collide, a large amount of vorticity is produced on small scales.
Since this is a turbulent effect, it increases as we increase the Reynolds number.
Indeed, the breakup of vortex sheets formed during vortex tube/ring collision is well documented
\citep{vanrees2012, Kerr:2013gd, PhysRevFluids.3.124702}.
For higher Reynolds numbers the flow becomes more turbulent and the non-viscous terms in equation \eqref{eq: enstrophy}
become more dominant.
It appears that the alignment of the fields is such that a net production of enstrophy is obtained.
In numerical vortex reconnection experiments \cite{Yao-Hussain-2020-883-A51-JFluidMech, Nguyen-Duong-2021-33-1-PhysFluids}
showed a similar behavior of enstrophy production during reconnection events.
With increasing Reynolds number, they too observe an increased enstrophy production.
The Coriolis contribution to the enstrophy evolution seems to dampen the production through the
term $(\uu\times\oo)\cdot\nab\times\oo$.

\section{Kinetic Helicity}

In the inviscid case the kinetic helicity (hereafter, simply ``helicity'') is conserved.
For a non-vanishing viscosity this will not be the case anymore.
However, for the present configuration, due to the symmetry of the configuration,
consisting of a vortex ring with positive and one with negative helicity, the volume-integrated
helicity is zero  and stays zero at all times.
Nevertheless, the existence of helicity in parts of the domain can influence  the relaxation.
Indeed, it has been shown that in the relaxation of magnetic braids,
not only the net helicity is important in constraining the dynamics, but also properties of the field line
mapping as well as the helicity-per-fieldline spectrum \citep{pontin2011, Yeates_Topology_2010, Yeates:2015dk}.
A way to detect the existence of a non-vanishing helicity density in the domain is to track
the evolution of the unsigned kinetic helicity as the integral over the
magnitude of the helicity density
\EQ \label{eq: Hmag}
\bar{H}_{\rm kin} = \int_{V} |(\oo + \OO)\cdot(\uu + \UU)|\ \dd V,
\EN
where $\nab\times\UU = \OO$.
Note that here we include the background vorticity and velocity, that is we calculate
the helicity in an inertial frame rather than the co-rotating frame.
This is for two reasons.
First, helicity conservation holds for the inertial frame but not in general for an accelerated frame.
Second, only in the inertial frame do we have properly braided vorticity lines in the initial state,
while in the corotating frame the kinetic helicity density vanishes everywhere at $t = 0$.

Furthermore, one has to note that unsigned kinetic helicity is not conserved, even under
conditions where the usual kinetic helicity is conserved.
For instance, if an initially straight untwisted vortex tube fixed between two parallel
plates is deformed by a rotation in the middle then the helicity is conserved as the left
and right hand twist in the tube cancel, but the unsigned helicity increases.
Nevertheless the unsigned helicity is always positive and can vanish only if the helicity
density vanishes everywhere in the domain.
The latter property is important for what follows, as it captures any non-zero helicity density in the domain.

We have run our simulations for different values of $\Rey$, and it is instructive to plot the results using time
units that are normalized by the viscous dissipation time scale
\EQ \label{eq: dissipation time}
\tau \approx L^2/\nu,
\EN
where $L$ is a typical length.
Here we take $L$ to be the (major) radius of the vortex rings which is
approximately $1$.

\begin{figure}\begin{center}
\includegraphics[width=0.95\columnwidth]{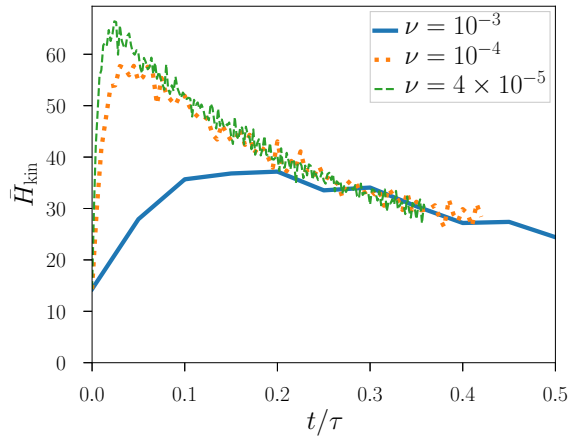}
\end{center}
\caption[]{
Evolution of the unsigned kinetic helicity for the relaxing vortex braid for different viscosities $\nu$
against normalized time (see equation \eqref{eq: dissipation time}).
}
\label{fig: Hmag_compare}\end{figure}

For our simulations we observe first a steep rise of $\bar{H}_{\rm kin}$ and then dissipation.
Since for $\nu = 10^{-3}$ we have $\tau = 1000$ this means that by time $1000$ we should observe a significant decrease in kinetic helicity.
Indeed, at time $t/\tau = 0.5$ we observe a drop by a factor of $e^{-0.5}$
compared to its peak value at early times (\Fig{fig: Hmag_compare}).

The initial rise happens at approximately $100$ code time units, independent of the viscosity,
which means that it is a non-viscous effect.
As the two initial vortex rings approach we observe an increase in kinetic helicity density until
the time of collision, which is approximately $100$ code time units.
This we attribute to vortex stretching.
After that we observe that the viscosity takes over and dissipates $\bar{H}_{\rm kin}$.
Note that for $\nu = 10^{-3}$, 100 code time units corresponds to $t/\tau=0.1$, while for $\nu = 10^{-4}$ the collision time is
$t/\tau=0.01$ in normalized times.

\subsection{Kinetic Helicity and Enstrophy}
\begin{figure}\begin{center}
\includegraphics[width=0.95\columnwidth]{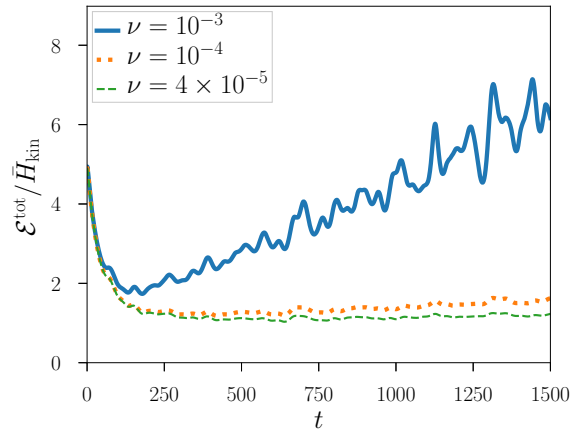}
\end{center}
\caption[]{
Evolution of the ratio of enstrophy with unsigned kinetic helicity for different viscosities $\nu$
against simulation time.
}
\label{fig: enstrophy_Hmag_t_compare}\end{figure}

The motivation to consider the relation between helicity and enstrophy comes from the magnetic case
where we know that the magnetic energy is limited from below by the magnetic helicity.
This is known as the realizability condition,
\citep{ArnoldHopf1974, BrandenbDoblerSubramanian2002AN, fluxRings10, knotsDecay11}. 
\EQ \label{eq: mineq}
\left|H_{\rm mag}  \right| \le \frac{2}{\lambda} E_{\rm mag} ,
\EN
where $\lambda$ is the smallest positive eigenvalue of the curl operator
in the domain \cite{ArnoldHopf1974, Arnold-TopologicalMethodsinHydrodynamics-2013}. 
The inequality is sharp, that is there exist fields for which the equality holds and these
are the eigenfields of the curl operator for the minimal eigenvalue $\pm \lambda$.
The corresponding condition for vorticity fields would involve the enstrophy
in the rest frame ${\cal E}^{\rm tot}$ and not the energy:
\EQ \label{eq: enstrineq}
\left|H_{\rm kin}  \right| \le \frac{1}{\lambda} {\cal E}^{\rm tot}.
\EN
Since the helicity is defined in the rest frame we also have to use the definition
of the enstrophy in the rest frame:
\EQ
{\cal E}^{\rm tot} = \int_V (\oo+\OO)^2\ \dd V.
\EN
In our case this inequality is not very useful since $H = 0$, which does not pose any lower bound on the enstrophy.
However, as shown in the Appendix (\ref{proofunsignedhelicity}) we can find an even stronger inequality:
\EQA \label{eq: ineq2}
\bar{H}_{\rm kin} & \le & \frac{1}{\lambda} \mathcal{E}^{\rm tot} \\
\Leftrightarrow \lambda  & \le &  \frac{\mathcal{E}^{\rm tot}}{\bar{H}_{\rm kin}} 
\ENA
The minimal $\lambda$ is not easy to determine for our domain, but one can approximate by using
the known minimal $\lambda = (2.405..)/R$ for a cylinder.
The largest cylinder we can fit into our domain has $R=5$, hence the $\lambda$ for our domain should be roughly  0.48.
The ratio $\mathcal{E}^{\rm tot}/\bar{H}_{\rm kin}$ is shown in \Fig{fig: enstrophy_Hmag_t_compare} and we clearly
see that the enstrophy is bounded from below by the unsigned kinetic helicity with a ratio $\ge 1$.

For the curve with the highest viscosity we find after the initial relaxation an increase of the
ratio $\mathcal{E}^{\rm tot}/\bar{H}_{\rm kin}$.
This is a result of our particular set-up.
As the viscous dissipation does not act on the fixed background vorticity field,
the evolution will eventually dissipate everything but the background field and for the latter the
ratio $\mathcal{E}^{\rm tot}/\bar{H}_{\rm kin}$ is infinite.
Hence after the first dynamic relaxation the ratio will eventually increase again also for the the other two curves.

\subsection{Kinetic Helicity and Kinetic Energy}

Next we study the relation between the integrated kinetic helicity and kinetic energy.
In order to be consistent with the helicity calculation we need to calculate the energy also in the rest frame, with velocity
$\uu + \UU$ and vorticity $\oo + \OO$.
With a much larger $\UU$ compared to $\uu$ (since $|\UU|\propto r \in [45,65] $),
the energy would be dominated by the background velocity. So in order to prevent that any change is obscured by the large background contribution, we compute the ``free'' kinetic energy instead, that is we subtract the energy of the background field
\EQ\label{eq: Ekin_free}
E_{\rm kin}^{\rm free} = \int_{V} \rho \left( \frac{1}{2} \uu^2 + \uu\cdot\UU \right) r \ \dd r\ \dd \theta\ \dd z.
\EN

Due to the Coriolis force,inertial waves are induced whose dominant frequency is determined by the
background rotation rate.
Combined with a background velocity $\UU$ that is large compared to $\uu$ we see large periodic
fluctuations of $E_{\rm kin}^{\rm free}$.
Therefore, to reveal the limiting behaviour we compute running means for our values $E_{\rm kin}^{\rm free}/\bar{H}_{\rm kin}$ over
$100$ time units.

\begin{figure}\begin{center}
\includegraphics[width=0.95\columnwidth]{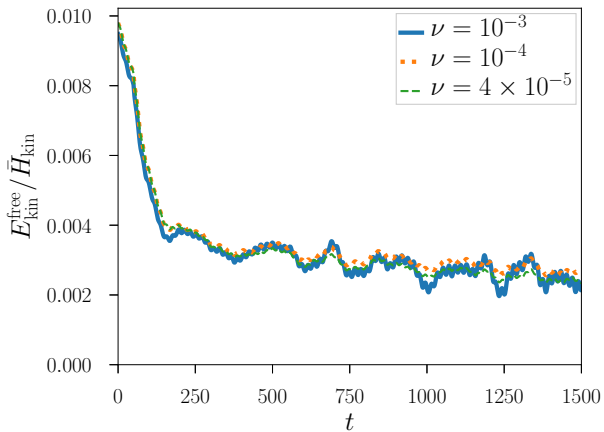}
\end{center}
\caption[]{
Evolution of the ratio of free kinetic energy with unsigned kinetic helicity for different viscosities $\nu$
against simulation time.
Here we use a running mean with window of $100$ time units to smooth out oscillatory behavior
induced by inertial waves.
}
\label{fig: ekinfree_Hmag_t_compare}\end{figure}

Although a strict lower limit for the kinetic energy is not known,
we observe that the ratio of the free kinetic energy and unsigned kinetic helicity tends asymptotically to a non-zero value
(\Fig{fig: ekinfree_Hmag_t_compare}) with the limit value of ca.\ 0.0025.
This is so striking that it leads us to conclude that there exists a lower limit for the
kinetic energy in the presence of unsigned kinetic helicity.

This finding is complementary to previous findings on helical turbulent flows in
rotating frames \citep{Teitelbaum-Mininni-2011-23-6-PhFluids, Teitelbaum-Mininni-2009-103-4-PRL}
where the authors studied the effect of net kinetic helicity and rotation
on the kinetic energy decay.
They find that, while helicity in a non-rotating frame does not affect the energy decay,
in a rotating frame, helicity poses restrictions leading to a slower decay.

\section{Field Topology}
\label{sec: field topology}

Our simulated configuration consists of two vortex rings.
However, we aim to compare our results to previous works using three pairs of such vortex rings
(e.g.\ \citet{Wilmot-Smith-Hornig-2009-696-1339-ApJ}).
Therefore, for the discussion in this section, we will make use of the periodicity in
the $z$-direction and construct such a braid by following vortex
streamlines over three periods.

\subsection{Simplification of the Topology}

\begin{figure}\begin{center}
\includegraphics[width=0.95\columnwidth]{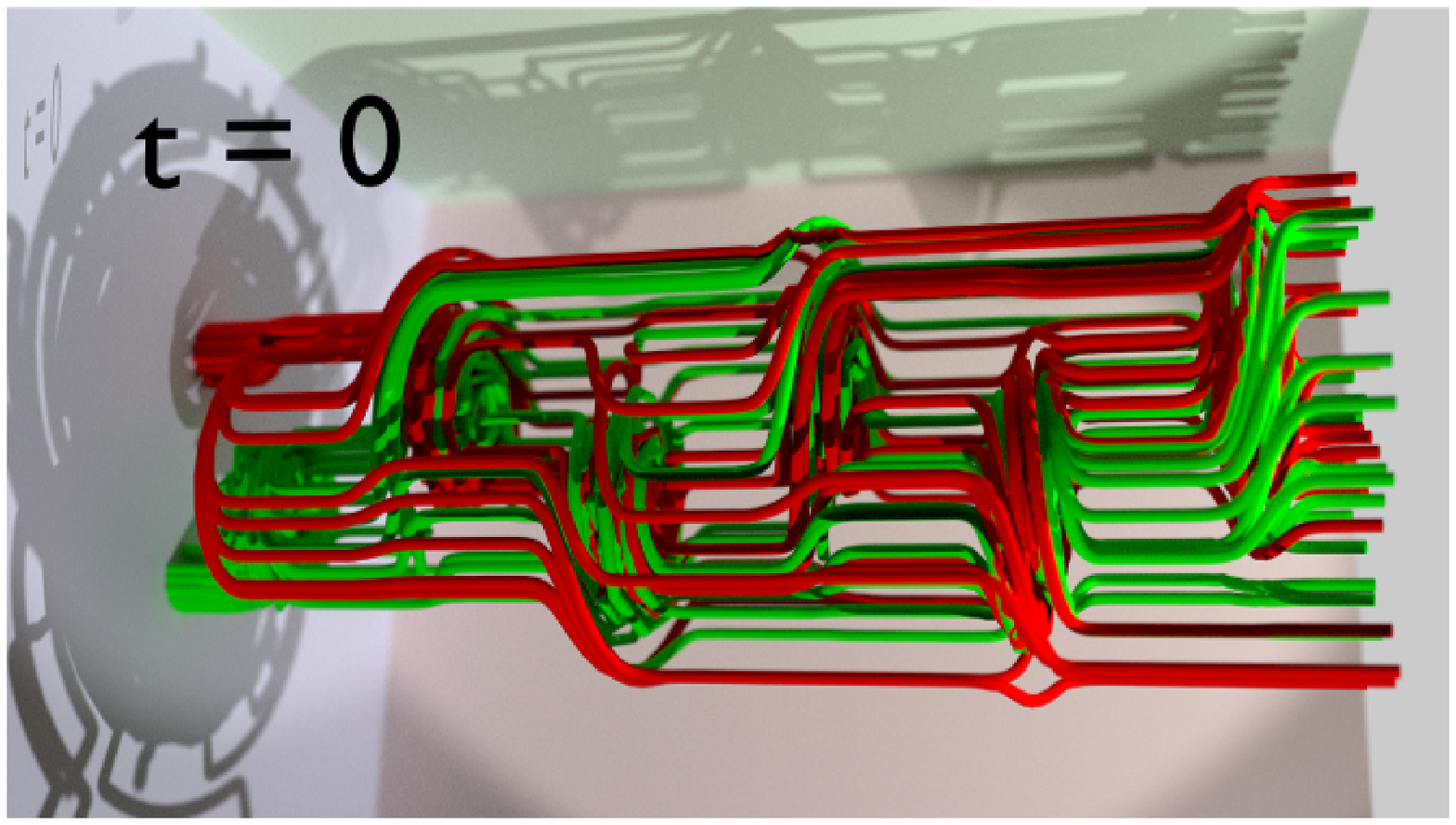} \\
\includegraphics[width=0.95\columnwidth]{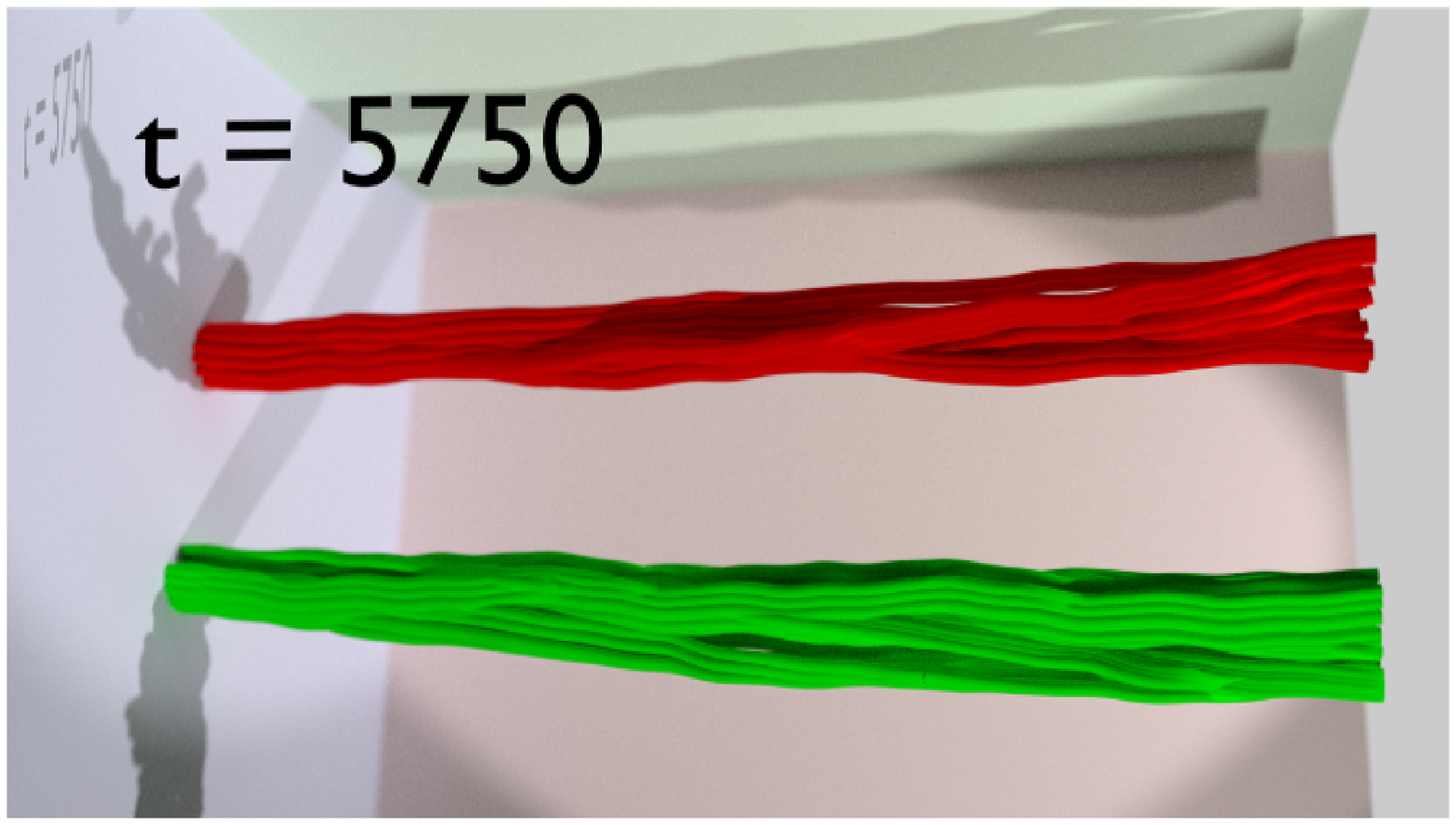}
\end{center}
\caption[]{
Vortex streamlines
in the rest frame (with background vorticity)
for the $\nu = 4\times 10^{-5}$ case at time $t = 0$ (upper panel)
and $t = 5750$ (lower panel).
In order to compare with previous simulations of magnetic braids we repeat our computational domain
in the $z$-direction (horizontal in these plots) three times.
The initial braid is largely unbraided and the final configuration consists of two separated
vortex tubes of opposite twist.
}
\label{fig: streamlines_t5750}\end{figure}

In order to analyze the changing topology of the vorticity field we integrate -- at each instant of time -- a set of vorticity field
lines starting from a fixed grid of starting points on the lower boundary ($z=-16$).
Naively we would expect the vortex field to simplify into a homogeneous field in the $z$-direction
(due to the net-zero helicity this should be the lowest-energy state).
However, as time progresses, and the field lines reconnect due to the finite viscosity, the topology of the field simplifies 
(\Fig{fig: streamlines_t5750}) not to an untwisted field, but -- similar to the magnetic case
\citep{pontin2011,Yeates_Topology_2010} -- into two large-scale vorticity
tubes containing twisted vortex lines, of opposite twist (swirl).
The fact that this final state mirrors closely the final state of the relaxation of a magnetic braid in a
plasma (with the same initial topology) suggests that some unifying underlying conservation principle is shared between the two systems.

\subsection{Field Line Helicity}

The above-mentioned simplification of the topology is effectively quantified/visualized by plotting the
\emph{kinetic field line helicity}, constructed as follows.
Due to the positive $z$-component of the vorticity, any field line starting at the
lower domain boundary will end at the top boundary.
This way we can find a one--to--one mapping between the boundaries.
For that we trace $256^2$ field lines starting at the lower boundary that are equally spaced in the radial direction
$r \in [49, 60]$ and the azimuthal direction $\theta \in [-0.1, 0.1]$.
We use those field lines to compute the kinetic field line helicity
\EQ \label{eq: curlyA}
\mathcal{A}(x_0, y_0) = \int_C \frac{((\uu + \UU)(x, y, z)\cdot(\oo(x, y, z)+\OO)}{(\oo(x, y, z)+\OO)_z} \ \dd z,
\EN
where $x(x_0, y_0, z)$ and $y(x_0, y_0, z)$ are the mapped points along the vorticity field line paths,
$C$ \citep{Yeates-Hornig-2011-18-102118-PhysPlasm}.
This measures the amount of winding \citep{Berger-1988-201-355-AA} of each field line
around all other field lines and gives us a
picture about the distribution of the helicity even in cases where its net value vanishes.

\begin{figure}\begin{center}
\includegraphics[width=0.95\columnwidth]{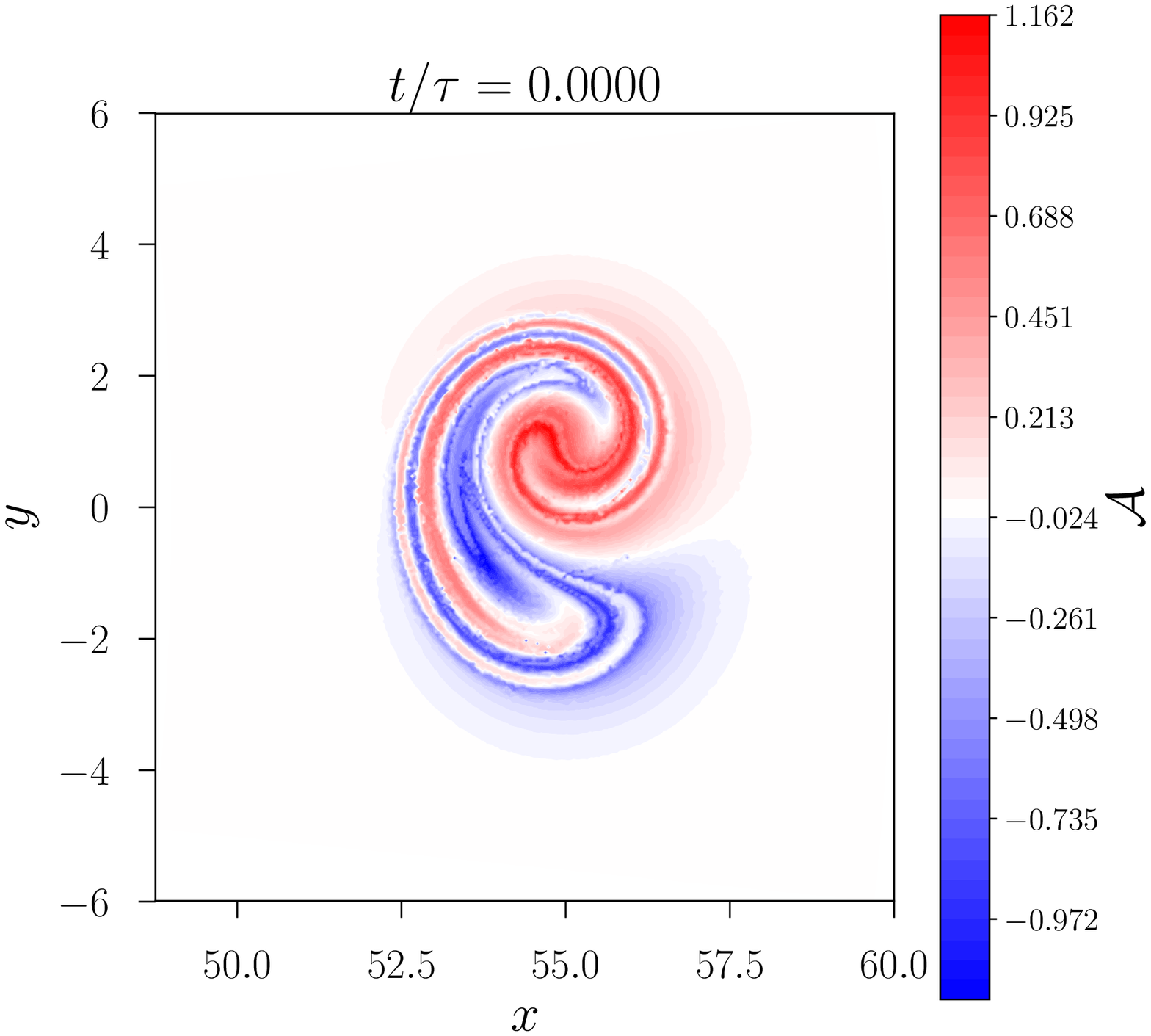} \\
\includegraphics[width=0.95\columnwidth]{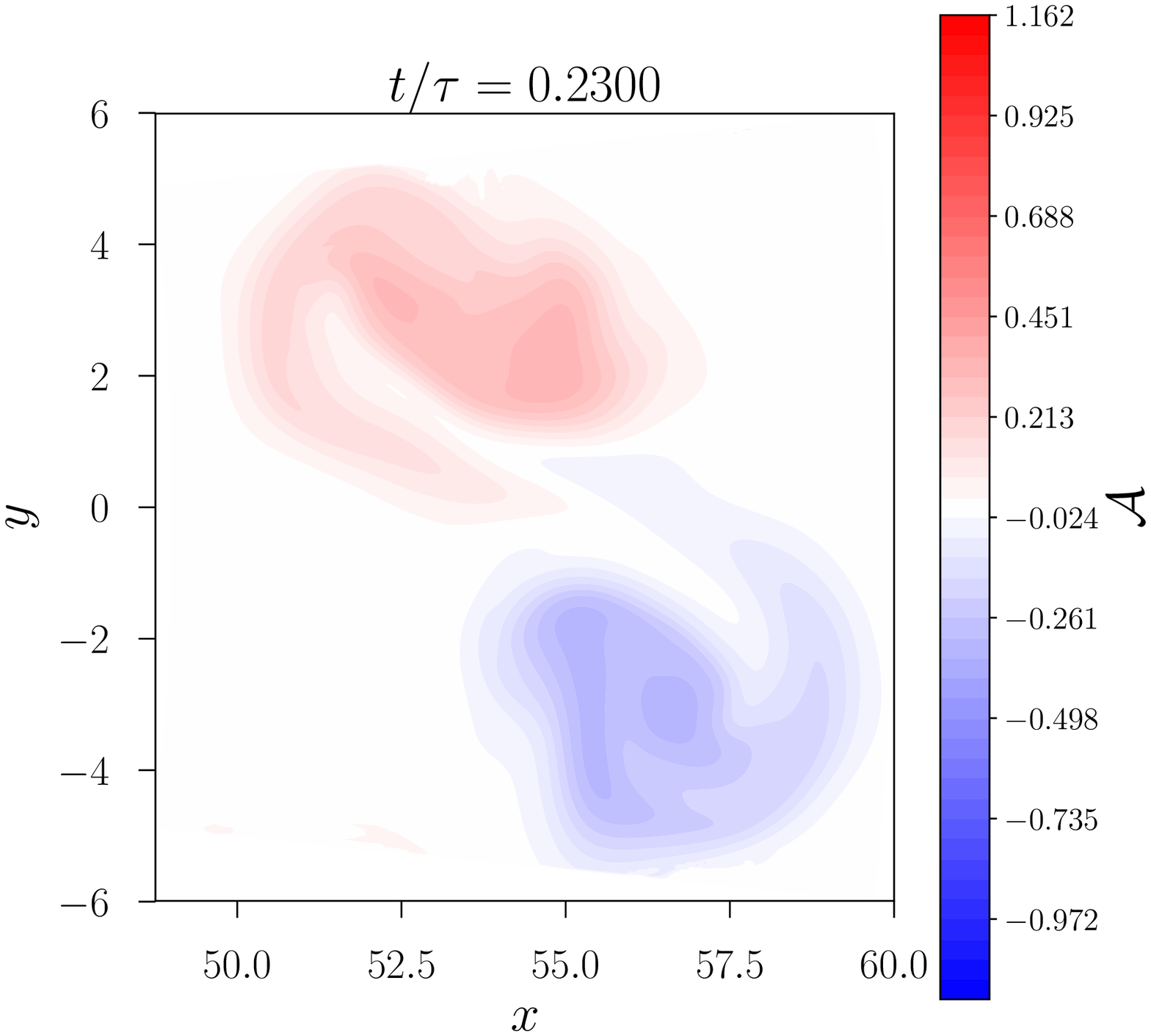}
\end{center}
\caption[]{
Field line helicity distribution
in the rest frame (with background vorticity)
of the vortex field at initial time (upper panel) and normalized time $t/\tau = 0.230$ (lower panel),
which corresponds to $t = 5750$ (see equation \eqref{eq: dissipation time}),
for the run with $\nu = 4\times 10^{-5}$.
In order to compare with previous simulations of magnetic braids we repeat our computational domain
in the $z$-direction three times.
Here we use $x = r\cos{(\theta)}$ and $y = r\sin{(\theta)}$ for the coordinates.
}
\label{fig: curlyA}\end{figure}

Since our vortex braid is highly tangled, the distribution of the field line helicity at
initial time shows some complexity at relatively small scales (\Fig{fig: curlyA}, upper panel).
As time progresses and the field lines reconnect, the distribution simplifies
greatly into two separate regions with opposite field line helicity (\Fig{fig: curlyA}, lower panel),
that correspond to the two flux tubes of opposite twist/swirl described above.

Indeed, \cite{Yeates-Hornig-2011-18-102118-PhysPlasm} showed that there is a connection between the reconnection
rate and the source term of the field line helicity, which for the hydrodynamic case takes the form
\EQ
\frac{\DD \mathcal{A}}{\DD t} = -\nu\int_C \frac{\nab\times\oo \cdot (\oo+\OO)}{|(\oo+\OO)|} \ \dd l,
\EN
where $l$ is the arc length along the vorticity line $C$.

\section{Conclusions}

We performed simulations of the relaxation of non-helical vortex braids in a cylindrical
wedge domain for a viscous fluid.
While the kinetic energy viscously decays, we observe an increase in the integrated
norm of the kinetic helicity density at dynamical times.
This increase is due to the reconnection of the vortex field lines at early times
and conincides with the time the flux rings that generate the braid collide.

The most striking finding of our study is that the unsigned kinetic helicity
appears to constrain the relaxation of the studied vortex braid.
Specifically, the ratio of the kinetic energy to unsigned helicity approaches a
non-zero value at late times that is independent of the viscosity.
This implies the presence of additional topological constraints on the hydrodynamic 
relaxation process, that may be related to those discovered recently for the magnetohydrodynamic system. 
In magnetohydrodynamics it is known that the presence of magnetic helicity imposes a
lower bound for the magnetic energy.
At the same time, we know from numerical experiments that topologically non-trivial
magnetic braids are not free to decay, even in the case of net-zero magnetic helicity.
The presence of additional topological constraints, such as preservation of the fixed point index or the field line helicity,
restrict the field's decay \citep{Yeates_Topology_2010, Yeates:2011gr, Yeates:2015dk}.
For the hydrodynamic case, such relations between the kinetic energy and kinetic
helicity are not known, but the present results strongly suggest their existence.

However, for the enstrophy we derived a lower bound in presence of unsigned kinetc helicity.
This relation is similar to the magnetohydrodynamic case, but with the enstrophy replacing
the energy.
Our simulations clearly confirm the validity of this analytical result
and we suggest that this relation should be taken into account when studying
the relaxation of hydrodynamical systems.

In addition to the above, we discovered another close parallel between the final
states of our vortex braid relaxation and magnetic braid relaxations.
Specifically, for the same braid topology, the two cases relax towards a topologically equivalent
final state, as revealed by plotting, e.g., the field line helicity (Figure \ref{fig: curlyA}).
The fact that the final states of these two very different relaxation processes
are analogous for the braid considered suggests that the constraints are likely
also related to one another, and the exploration of these topological 
constraints in the hydrodynamic system will be an important area of future study.

\section*{Acknowledgements}

SC, DP and GH acknowledge financial support from the UK's
STFC (grant number ST/N000714/1 and ST/S000267/1).
For the plots we made use of the Matplotlib library for Python \citep{Hunter:2007}
and BlenDaViz\footnote{github.com/SimonCan/BlenDaViz}.

\section*{Data Availability}

Raw data were generated using HPC facilities.
Derived data supporting the findings of this study are available from
the corresponding author upon reasonable request.

\appendix

\section{Construction of Vortex Tubes}
\label{sec: construction of vortex tubes}

In our simulations, the variables that are solved for are not the vorticity $\oo$,
but the velocity $\uu$.
So, we need to specify our initial conditions in terms of $\uu$ with $\nab\times\uu = \oo$.
Yet, for a given vorticity $\oo$ we can find different velocities $\uu$ such
that $\nab\times\uu = \oo$, similar to the gauge freedom for the magnetic
vector potential $\AAA$ with the magnetic field $\BB = \nab\times\AAA$.
However, it is not desirable to use the expression for the vector potential
from \citet{Wilmot-Smith-Hornig-2009-696-1339-ApJ} for our velocity field $\uu$ as it is not divergence-free,
leading to unwanted compression.

In order to construct a divergence-free flow field we use the solutions of the Biot-Savart integral
for a singular vortex ring
(see \citet{jackson2007classical}, section 5.5).
We then construct the vortex ring from a sum (integral) of infinitely
many infinitesimally thin vortex rings.
For that we compute a potential $C$ such that $\uu = \alpha\nab\times(C\ee_\theta)$
which results in a divergence-free velocity field by construction.

We first construct the potential $C_0$ for a single vortex ring in a coordinate system
with origin at the ring's center.
In a later step we will shift (transform) this potential to its actual position.
Our coordinates here are $(r_0, \theta_0, z_0)$.
Here, the potential $C_0(r_0, \theta_0, z_0)$ is the double integral
\EQ \label{eq: C0 potential}
C_0(r_0, \theta_0, z_0) = \int_{-8}^{8}\dd z_0' \int_{0}^{5} \dd r_0' \sqrt{2}r_0'
e^{-r_0'^2/2 - z_0'^2/4} \Psi,
\EN
with
\EQ
\Psi = \frac{r_0' \left( (\kappa^2+2)K(\kappa)-2E(\kappa) \right)}
{\pi\sqrt{r_0'^2 + r_0^2 + (z_0 - z_0')^2 + 2r_0'r_0)}\kappa^2},
\EN
with the complete elliptical integral of the first kind $K(\kappa)$, complete elliptical integral $E(\kappa)$ and
\EQ
\kappa = 2\sqrt{\frac{r_0'r_0}{r_0'^2 + r_0^2 + (z_0 - z_0')^2 + 2r_0'r_0}}.
\EN
We choose to integrate in $z_0'$ from $-8$ to $8$ and in $r'_0$ from $0$ to $5$,
as beyond those integration intervals the integrand is sufficiently small to
be neglected from the integration.
For more details about this construction see \citet{jackson2007classical}, section 5.5.

This gives us the vector potential $C_0(r_0, \theta_0, z_0){\bf e}_{\theta_0}$ in the centered
coordinate system $(r_0, \theta_0, z_0)$.
In order to construct the braid we make a coordinate transformation so that our ring is centered at $r=R_0, \theta=\Theta$.
The coordinates transform according to
\EQA
r_0 & = & \sqrt{r^2 + R_0^2 - 2r R_0\cos(\theta-\Theta)} \\
\theta_0 & = & \arctan(\sin(\theta-\Theta)/\cos(\theta-\Theta) - R_0/r) \\
z_0 & = & z,
\ENA
while the vector potential transforms as
\EQA
C_r(r, \theta, z) & = & \frac{-C_0(r_0, \theta_0, z_0) R_0 \sin(\theta-\Theta)}{\sqrt{-2R_0 r\cos(\theta-\Theta) + R_0^2 + r^2}} \\
C_\theta(r, \theta, z) & = & \frac{C_0(r_0, \theta_0, z_0) (r - R_0\cos(\theta-\Theta))}{\sqrt{-2R_0 r\cos(\theta-\Theta) + R_0^2 + r^2}}.
\ENA
After this transformation we apply the curl operator in the wedge domain and obtain the initial velocity
in the wedge domain.
The resulting vortex rings have a minor and major radius of ca.\ $1$.

\section{Inequality for the Unsigned Helicity} \label{proofunsignedhelicity}

The relation between unsigned kinetic helicity and enstrophy is derived using
the Poincar\'e inequality 
\EQ
\left(\int_V \|\uu\|^2\ \dd^3x\right)^{1/2} \le \frac{1}{\lambda} \left(\int_V \|\oo\|^2\ \dd^3x\right)^{1/2},
\EN
which holds for every differentiable field, such that $\oo = \nab\times\uu$.
Here $\lambda$ is the smallest positive eigenvalue of the curl operator.

We can now apply the Cauchy-Schwarz inequality to the unsigned kinetic helicity density to obtain 
\EQ
\int_V \left| {\oo \cdot \uu}\right| \, \dd^3x  \le \int_V \|\oo\| \|\uu\|\ \dd^3x,
\EN
and apply it a second time to the integral ($L^2$-norm version)
\EQ
\int_V \left| {\oo \cdot \uu}\right| \, \dd^3x \le
\left(\int_V \|\oo\|^2\ \dd^3x\right)^{1/2} \left(\int_V \|\uu\|^2\ \dd^3x\right)^{1/2}.
\EN
Eventually we use the Poincar\'e inequality to obtain
\EQ
\int_V \left| {\oo \cdot \uu}\right| \, \dd^3x \le
\frac{1}{\lambda} \int_V \|\oo\|^2\ \dd^3x.
\EN
This leaves us with the inequality
\EQ
\bar{H}_{\rm kin} \le \frac{1}{\lambda}\mathcal{E}.
\EN

\bibliography{references}

\end{document}